\documentclass[conference]{IEEEtran}
\IEEEoverridecommandlockouts
\usepackage{adjustbox}
\usepackage{soul}
\usepackage{amsmath}
\usepackage{float}
\usepackage{caption}
\usepackage{tabularx}
\usepackage[table,xcdraw]{xcolor}
\usepackage{pifont}
\usepackage{changepage}
\usepackage{gensymb}
\usepackage{fontawesome}
\usepackage{multirow} 
\usepackage[edges]{forest}
\usepackage{makecell}
\usepackage{amssymb}
\usepackage{booktabs}
\usepackage{tcolorbox}
\usepackage{tabto}
\usepackage{tikz}
\newcommand{\circled}[1]{\tikz[baseline=(char.base)]{
    \node[shape=circle,draw,inner sep=0.5pt] (char) {\scriptsize #1};}}
\usepackage{graphicx}
\usepackage{subcaption}
\usepackage{verbatim}

\usepackage{algorithmic}
\usepackage{diagbox}
\usepackage{tikz}
\usetikzlibrary{shapes.geometric}
\usetikzlibrary{positioning}
\usepackage{tabularx}
\usepackage{ragged2e}
\usepackage{lipsum} 
\usepackage{amsmath}
\usepackage{amssymb} 
\begin{document}
\bstctlcite{IEEEexample:BSTcontrol}




\title{Quality over Quantity: Value-Driven Distributed Congestion Control for the Collective Perception Service}

\author{
    Tengfei Lyu$^{1}$, Florian A. Schiegg$^{2}$, Md Noor-A-Rahim$^{1}$, Dirk Pesch$^{1}$, Aisling O'Driscoll$^{1}$ \\
    $^{1}$nasc Research, School of Computer Science \& IT, University College Cork, Ireland.\\
    $^{2}$Corporate Research, Robert Bosch GmbH, Germany. \\
    Email: t.lyu@cs.ucc.ie$^{1}$, \{md.noorarahim, dirk.pesch, aisling.odriscoll\}@ucc.ie$^{1}$, \\ florian.schiegg@de.bosch.com$^{2}$
}

\maketitle

\begin{abstract}
The Collective Perception Service (CPS) enables Intelligent Transport System Stations (ITS-S') to exchange sensed information about surrounding objects and regions. However, the frequent transmission of Collective Perception Messages (CPMs), together with their highly variable size, can lead to severe channel congestion in dense environments. Existing Facilities layer Decentralised Congestion Control (DCC FAC) mechanisms improve upon content-agnostic schemes by prioritising higher-value objects and regions within the allocated Facilities layer bit budget. Nevertheless, current \textit{quantity} based content selection mechanisms remain limited in heterogeneous environments, where the amount of perception data and the distribution of it's Value of Information (VoI) scores can vary significantly across ITS-S' and over time. This paper proposes a DCC FAC \textit{quality} content selector based on adaptive VoI thresholding. 
This threshold ensures that limited radio resources are utilised in a more meaningful way to better prioritise the transmission of high VoI perception content. The proposed method is evaluated against the ETSI DCC FAC \textit{quantity} content selector in a controlled CPM/VoI simulation environment under homogeneous and heterogeneous network knowledge, and under low, medium, and high channel loads. The results show that selection based on the adaptive VoI threshold consistently retains higher network transmitted VoI, especially under heterogeneous perception knowledge.
\end{abstract}


\section{Introduction}
The Collective Perception Service as specified by the European Telecommunications Standards Institute (ETSI)~\cite{etsiintelligent}, improves the awareness of vehicles and infrastructure by allowing them to share information on sensed obstacles, other road users and their perceived regions. This shared view can support safer driving, smoother traffic flow, and better control decisions. However, it has been shown in~\cite{xhoxhi2023first} that CPMs incur high radio resource demands due to their frequent transmission and highly variable packet sizes, especially when optional fields are enabled. At the same time, real-world tests have also shown that CPM traffic can become very high in dense scenarios, leading to serious channel congestion~\cite{figueiredo2024enhancing}. Given limited dedicated ITS spectrum in the 5.9~GHz band, this motivates the need for congestion control measures that ensure prioritised transmission of highly relevant CPM perception content i.e. objects and regions, while maintaining channel stability.

To address this, ETSI defined the legacy \textit{Decentralised Congestion Control (DCC)} architecture~\cite{etsiintelligent2015}, in which congestion-control functions can operate across multiple layers of the ITS protocol stack. The most established mechanisms are located at the \textit{Access} layer, where channel load is controlled through transmission power, transmission rate, or data-rate adaptation based on the estimated Channel Busy Ratio (CBR). Such content-agnostic DCC mechanisms have been well studied and are known to have recognised limitations~\cite{bansal2013limeric,bansal2013achieving,bansal2013embarc,lorenzen2017swerc}. More recently, ETSI Release~2 has highlighted the need for \textit{Facilities} layer resource management (\textit{DCC FAC}) that allows services to adapt their message generation to the currently available communication resources~\cite{etsiintelligent2021DCC,etsiintelligent2022,etsiintelligentRRM}. In this context, ETSI DCC FAC supports Value of Information (VoI) based content selection within the CPS, respecting specified resource limits. Specifically, in a recent technical study~\cite{etsiintelligentRRM}, a \textit{quantity} based ETSI DCC FAC content selector is considered. Candidate objects and regions are ranked in descending VoI order and included for CPM assembly until the allocated Facilities layer bit budget is exhausted. While this performs better than traditional content-agnostic DCC mechanisms, it implicitly assumes homogeneity, i.e that all stations sense broadly similar numbers of objects/regions with comparable distribution of VoI scores across Intelligent Transportation System Stations (ITS-S') or over time. 

However heterogeneous network knowledge is more commonplace. Between decentralised ITS-S', the distribution of VoI is inherently heterogeneous across both space and time. Differences in sensing capabilities, viewpoints, occlusion conditions, traffic context, and prior knowledge lead to significant variation in perception content value across ITS-S'. The implementation of the VoI function will also impact the distribution of the VoI values over time which requires further study. Consequently, equal resource allocation does not guarantee the dissemination of equally valuable CPM content.

These observations highlight the need for DCC mechanisms that not only mitigate channel congestion but also explicitly prioritise resource allocation for ITS-S' with high-VoI content, particularly under heterogeneous network knowledge. This paper addresses this with the following contributions:


\begin{itemize}
    \item A DCC FAC \textit{quality} content selector is proposed based on adaptive VoI thresholding. Starting from the provided Facilities layer resource budget, the proposed method dynamically adjusts the admissible VoI threshold so that content inclusion becomes more selective under congestion and less selective when channel conditions allow.

    \item A comparative evaluation is conducted against the ETSI DCC FAC \textit{quantity} content selector in a controlled CPM/VoI simulation environment with homogeneous and heterogeneous network knowledge.
\end{itemize}

The results show that the proposed method improves the overall \textit{network transmitted VoI} in all settings, with moderate gains under homogeneous knowledge and substantially larger gains under heterogeneous knowledge. The remainder of this paper is organised as follows. Section~\ref{relevant_background} reviews the relevant state-of-the-art DCC mechanisms, including recent value-based approaches. Section~\ref{proposed} motivates and presents the proposed value-based DCC \textit{quality} selector with adaptive VoI thresholding. Section~\ref{simulation_scenario} describes the simulation environment and utilised models, with Section~\ref{evaluation} highlighting the performance benefits of the proposed approach. Section~\ref{conclusions} concludes the paper and outlines directions for future work.

\section{State of the Art in Decentralised Congestion Control}
\label{relevant_background}
In this section, we review the most pertinent DCC mechanisms that form the benchmark for our study, both in standards and in the academic literature, and motivate the need for advanced value-based schemes.

\subsection{ETSI DCC Access}
Historically, the ETSI DCC \textit{Access layer} has regulated channel load through three main approaches~\cite{etsi2018intelligent}: Transmission Power Control (TPC), Transmission Rate Control (TRC), and Transmission Data Rate Control (TDC).
TRC, which is most commonly employed, adapts the temporal rate at which messages are sent, typically by increasing the minimum interval between consecutive transmissions under congestion.
Two variants exist: \textit{Reactive DCC (R-DCC)} and \textit{Adaptive DCC (A-DCC)}~\cite{etsi2018intelligent}. R-DCC follows a rule-based design in which measured CBR ranges are mapped to different minimum inter-transmission intervals, i.e. higher measured CBR leads to a more restrictive message rate. In response to known R-DCC limitations, A-DCC employs a continuous linear-feedback controller derived from \textit{LIMERIC}~\cite{bansal2013limeric}, and updates the allowed channel occupancy of each ITS-S according to the difference between the measured and target CBR~\cite{bansal2013limeric}. Each ITS-S $j$ maintains a $\delta_j(t)\in[0,1]$, representing the fraction of the channel resources that it is allowed to use at time $t$. Every $200~\mathrm{ms}$, A-DCC updates $\delta_j(t)$ as:
\begin{equation}
\label{eq:A-DCC}
  \delta_j(t) = (1-\alpha)\,\delta_j(t-1) + \beta \bigl(\mathrm{CBR}_{\mathrm{target}} - \mathrm{CBR}_{\mathrm{ITS-S}}(t-1)\bigr)
\end{equation}
where $\alpha \in (0,1)$ is a smoothing factor, $\beta>0$ is the adaptation gain, and $\mathrm{CBR}_{\mathrm{target}}$ is the target CBR. If the measured CBR exceeds the target, $\delta_j(t)$ is reduced; otherwise, it increases. The resulting channel-occupancy allowance is then enforced through the corresponding inter-transmission interval. 

\subsection{ETSI DCC FAC - Reference Architecture}
\label{etsi_dcc_fac_background}
\begin{figure}[t]
        \centering
        \includegraphics[width=\columnwidth]{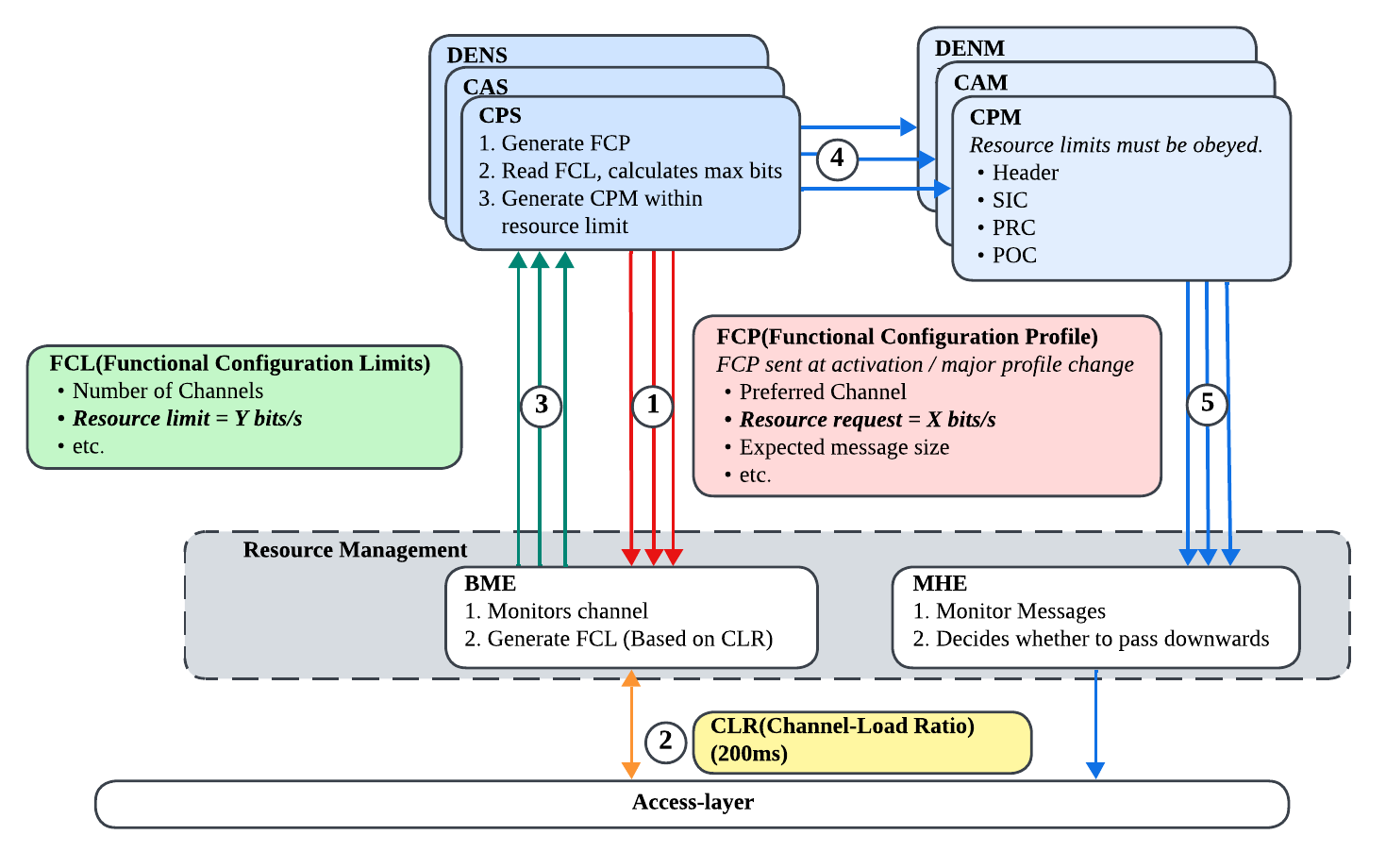}
        \caption{The ETSI DCC Facilities layer (DCC-FAC) Architecture defined in~\cite{etsiintelligent2021DCC, etsiintelligent2022} including the recent extensions for Resource Management~\cite{etsiintelligentRRM}.}
        \label{fig:etsi_dcc_fac}
        \vspace{-4mm}
\end{figure}
Shaping traffic only at the Access layer is insufficient when multiple ITS services coexist and compete for the same channel resources~\cite{khan2020enhancing,khan2019vehicle}, nor does it account for the \textit{actual service payload}. For this reason, ETSI introduced Decentralised Congestion Control at the \textit{Facilities layer} \textit{(DCC FAC)}, which operates closer to the services and controls how much traffic each service is allowed to inject into the lower layers. The first ETSI specification of the \textit{DCC FAC} entity was given in \textit{TS 103 141 V2.1.1}~\cite{etsiintelligent2021DCC}. This was followed by \textit{TS 103 141 V2.2.1}~\cite{etsiintelligent2022}, which placed the functionality within the Release~2 MCO/FAC framework and explicitly described the operation of the \textit{Bandwidth Management Entity} (BME), including the generation of \textit{Functional Configuration Profiles (FCPs)}, the use of lower-layer status information such as the \textit{Channel Load Ratio (CLR)}, and the derivation of \textit{Functional Configuration Limits (FCLs)} for each service~\cite{etsiintelligent2022}. More recently, ETSI proposed a \textit{Resource Management} component in \textit{TR 104 073}~\cite{etsiintelligentRRM} including the \textit{BME} and the \textit{Message Handling Entity} (MHE). Fig.~\ref{fig:etsi_dcc_fac} illustrates this updated architecture.

Individual ITS services such as the CPS, Cooperative Awareness Service (CAS), and Decentralised Environmental Notification Service (DENS) generate \textit{FCPs}, which describe their preferred channel, their requested resources in bit/s, and their expected message characteristics (label~\circled{1} in Fig.~\ref{fig:etsi_dcc_fac}). The FAC layer \textit{BME} continuously monitors the access layer channel load deriving a \textit{CLR} every $200$~ms (label~\circled{2}). Based on the current CLR and the set of FCPs, the BME computes the available communication resources and generates per-service \textit{FCLs} (label~\circled{3}). These FCLs translate the internally computed control state into parameters that are \textit{understandable} by the V2X services, such as a per-channel resource limit in \textit{bit/s} or an equivalent \textit{message generation interval}. Each service then uses its FCL as the input to its own message-generation process (label~\circled{4}). The resulting messages are passed to the \textit{MHE}, which checks whether the observed traffic complies with the current FCLs and forwards only compliant messages to the lower layers, while non-compliant messages are delayed or suppressed according to the Resource Management policy (label~\circled{5}).

\subsection{Value based DCC - Standards \& Literature}
\label{subsec:valuebaseddcc}
In contrast to the significant body of literature studying Access layer congestion control, there are limited studies evaluating \textit{value-based} DCC FAC mechanisms. A first step is provided by the ETSI DCC FAC \textit{\textbf{quantity}} content selector~\cite{etsiintelligent, etsiintelligent2021DCC, etsiintelligent2022}. Each perceived object or region is assumed to have a VoI score, with the function to derive these yet to be defined. Once the CPS receives its current FCL, it converts the corresponding resource allowance into a per-message bit budget, ranks content in descending VoI and inserts it into the CPM  until the allocated Facilities layer bit budget is exhausted. Hence, the ETSI DCC FAC \textit{quantity} content selector introduces a basic form of content awareness, since higher-VoI objects and regions are preferred over lower-VoI ones. However, it does not preferentially allow for the transmission of higher VoI content across ITS-S'.

A more explicit value-based approach was proposed by Wolff et al. in~\cite{wolff2025Uncertainty}. Each ITS-S is assigned a resource budget for the next duty cycle according to the VoI contribution of the objects it transmitted during the previous cycle, relative to the total VoI observed in received CPMs. A fixed VoI threshold is first applied so that only sufficiently valuable objects contribute to the budget calculation, and the resulting budget is then used for subsequent transmission decisions. This is an important step beyond the ETSI DCC FAC \textit{quantity} content selector because it not only prioritises objects within a given CPM budget, but also attempts to redistribute channel resources according to content value. Nevertheless, several limitations remain. First, the fixed VoI threshold cannot adapt its selectivity to changing channel conditions. Second, the mechanism is sensitive to parameter choices and smoothing-window effects. Third, the budget update introduces a temporal lag: an ITS-S that contributed high-VoI objects in one duty cycle may not necessarily observe equally valuable objects in the next, which can lead to temporary over-allocation of resources and inefficient channel use.

Most recently, Sepulcre et al.~\cite{sepulcre2026demand} adapted a \textit{weighted} LIMERIC based approach~\cite{bansal2013achieving} operating in the BME, in which ITS-S' are assigned different $\beta$ values so that their allowable transmission rates become proportional to predefined weights. The premise is that rather than attempting equal resource allocation amongst ITS-S', nodes that have higher communication demands (on the assumption of better sensing capabilities) are assigned larger $\beta$ values and therefore converge to larger resource shares. This is an important step towards a VoI based \textit{quality} controller at the Facilities layer BME, but it also highlights a limitation that is particularly relevant for collective perception: in dynamic traffic environments, a vehicle or ITS-S with enhanced sensed capabilities does not necessarily observe a large number of valuable objects consistently over time. Due to mobility, occlusion, and scene evolution, prioritising resources only at the node level may still allocate additional resources when the actual sensed content set is temporarily of limited value, thereby potentially leading to inefficient resource usage. It is preferable to prioritise resource usage based on the observed distribution of VoI values, not at a node level. 

\section{Proposed Facilities Layer DCC VoI Quality Content Selector}
\label{proposed}
\begin{figure*}[t]
  \centering
  \begin{subfigure}[b]{0.31\textwidth}
    \centering
    \includegraphics[width=\textwidth]{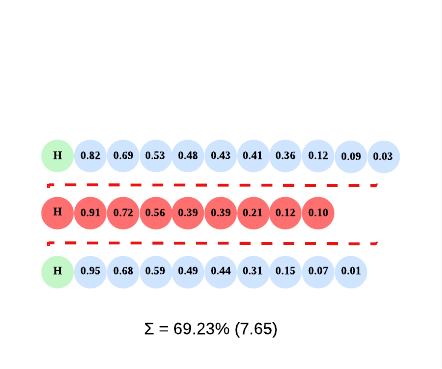}
    {\captionsetup{font=footnotesize}
    \caption{Traditional content-agnostic control}
    \label{fig:homo_load}}
  \end{subfigure}
  \hfill
  \begin{subfigure}[b]{0.31\textwidth}
    \centering
    \includegraphics[width=\textwidth]{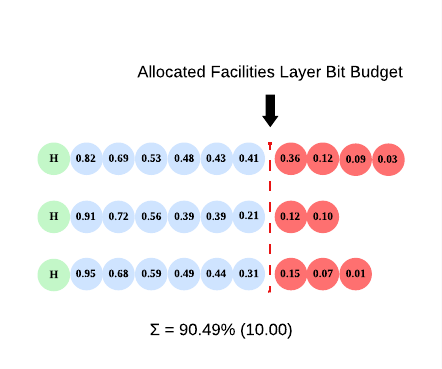}
    {\captionsetup{font=footnotesize}
    \caption{ETSI DCC FAC \textit{quantity} content selector}
    \label{fig:homo_fac}}
  \end{subfigure}
  \hfill
  \begin{subfigure}[b]{0.31\textwidth}
    \centering
    \includegraphics[width=\textwidth]{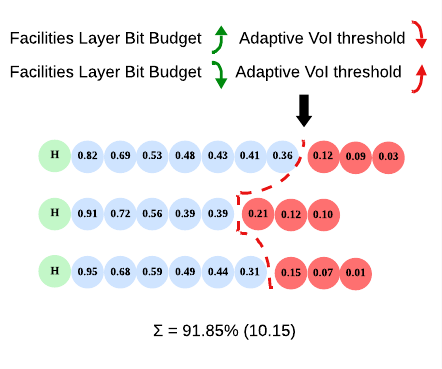}
    {\captionsetup{font=footnotesize}
    \caption{Proposed DCC FAC \textit{quality} content selector}
    \label{fig:homo_voi}}
  \end{subfigure}

  \caption{Network Transmitted VoI ($\sum$) of three DCC approaches under \textit{homogeneous} network knowledge, where the number of perceived objects/regions and the overall VoI distribution are statistically similar across ITS-S'.}
  \label{fig:homo_overview}
  \vspace{-4mm}
\end{figure*}

\begin{figure*}[t]
  \centering
  \begin{subfigure}[b]{0.46\textwidth}
    \centering
    \includegraphics[width=\textwidth]{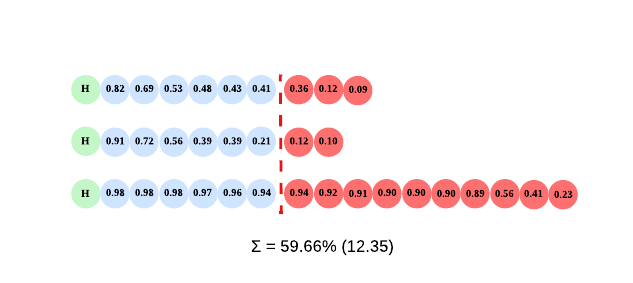}
    {\captionsetup{font=footnotesize}
    \caption{ETSI DCC FAC \textit{quantity} content selector}
    \label{fig:hete_fac}}
  \end{subfigure}
  \hfill
  \begin{subfigure}[b]{0.46\textwidth}
    \centering
    \includegraphics[width=\textwidth]{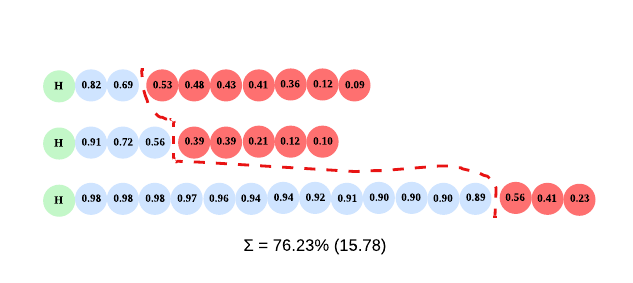}
    {\captionsetup{font=footnotesize}
    \caption{Proposed DCC FAC \textit{quality} content selector}
    \label{fig:hete_voi}}
  \end{subfigure}

  \caption{Network Transmitted VoI ($\sum$) of the quantity-based and proposed \textit{quality} content DCC content selectors under \textit{heterogeneous} network knowledge, where the number of perceived objects/regions and the distribution of VoI values may differ across ITS-S'.}
  \vspace{-4mm}
  \label{fig:hete_overview}
\end{figure*}

This paper proposes a value based \textit{\textbf{quality}} content selector that improves how the available service layer CPM budget is used under congestion by prioritising higher-VoI content through an adaptive VoI threshold. Unlike a \textit{quantity} based content selector, the proposed method does not simply try to include as much high-ranked content as possible under the current allocated Facilities layer bit budget, i.e. striving for \textit{equal} channel access across all ITS-S', but rather it optimises for the network transmitted VoI, i.e. maximising the collective perceptual awareness.  

\subsection{Value-driven \textit{Quality} Congestion Control}
The proposed value based \textit{quality} content selector adapts how selective an ITS-S is when selecting content to include in the CPM. Only content (objects or regions) whose value is sufficiently high remains eligible for transmission under congested channel conditions. To motivate this, consider Fig.~\ref{fig:homo_overview}, which provides an illustrative comparison of three DCC approaches on three ITS-S' at a single time point. Each horizontal line corresponds to one CPM generation event at time $t$ on a given ITS-S. The green circle (\textit{H}) represents the fixed CPM overhead, blue circles denote an object or region included in the CPM, and the number inside each circle indicates the content's VoI, with contents ordered from high to low. Red circles represent content that is excluded as it cannot be transmitted under the currently allocated Facilities layer bit budget. 

In Fig.~\ref{fig:homo_load}, a content-agnostic control approach such as A-DCC may suppress an entire CPM when the message exceeds the admissible load, even if it contains some high-VoI content. This illustrates the limitation of regulating congestion solely at the message level. Although such an approach can reduce channel load, it may discard useful high value perceptual content together with less useful objects and regions. In Fig.~\ref{fig:homo_fac}, the ETSI DCC FAC \textit{quantity} selector improves upon this by explicitly considering content value. It ranks and selects content in descending VoI order until the CPM budget is exhausted. This avoids dropping the full message when at least part of the content can still be transmitted. However, it has no mechanism to adapt its selectivity when a perceived content set on ITS-S $i$ becomes dense with low-value perceptual information, while high value content on ITS-S $j$ might be omitted. In contrast, the proposed \textit{quality} selector in Fig.~\ref{fig:homo_voi} introduces a stronger form of value-awareness. It applies an adaptive minimum VoI threshold so that content with insufficient value is discounted before packet assembly. Therefore, the proposed method is designed not only to rank content by value, but also to regulate which content should be prioritised for transmission as channel congestion increases or decreases. In this way, the VoI selector threshold becomes more restrictive, i.e. goes up when the allocated Facilities layer bit budget is insufficient for the content set, and becomes less restrictive, i.e. goes down when channel conditions are more favourable. 

\subsection{Homogenous vs heterogeneous Network Knowledge}
The benefits of the \textit{quality} content selector become particularly evident when considered under \textbf{\textit{homogeneous}} vs \textbf{\textit{heterogeneous}} network knowledge. In this study, homogeneity and heterogeneity are contrasted along two dimensions: the number of perceived objects/regions and the distribution of VoI scores. Each dimension may be considered across different ITS-S' or within a single ITS-S over time, although in this study, the emphasis is placed on comparisons across different ITS-S' which is shown in Figs.~\ref{fig:homo_overview} and~\ref{fig:hete_overview}. Homogeneity does not mean that the VoI scores of all content are similar within a single ITS-S. 

\begin{itemize}
    \item For the \textit{\textbf{number of perceived objects}}, \textit{\textbf{homogeneous}} means that a similar number of objects is perceived across different ITS-S', whereas \textit{\textbf{heterogeneous}} means that the number of perceived objects varies between different ITS-S'. The former may arise in relatively simple environments, such as motorway scenarios, where neighbouring ITS-S' observe broadly similar scenes, whereas the latter is more likely in complex environments, such as busy junctions, where occlusions, differing fields of view, and local traffic density may cause some ITS-S' to perceive substantially more content than others.
    \item For the \textit{\textbf{distribution of VoI scores}}, \textit{\textbf{homogeneous}} means that the content VoI scores are similar across different ITS-S', whereas \textit{\textbf{heterogeneous}} means that the distribution of object VoI scores varies between different ITS-S'. This is largely depends on how the VoI for the CPM content is determined. For example, in the VoI functions specified in~\cite{wolff2025Uncertainty} and~\cite{lyu2025accuracy}, the value of an object is not determined by a single factor, but by the interaction between multiple components, including its perceived accuracy and its relevance to neighbouring ITS-S'. In~\cite{figueiredo5389372improving} newly sensed objects are assigned high VoI scores. An object may have lower VoI if it is poorly perceived, but also if it is already well known to surrounding nodes. Conversely, an object that is both accurately perceived and not yet widely known may become highly valuable. As vehicles move, content that was previously unimportant may suddenly become highly relevant, while other content may lose value. Thus heterogeneous VoI distributions are much more likely in practice as homogeneous ones will typically only arise when neighbouring ITS-S' observe similar objects under comparable sensing conditions and maintain a similar level of prior awareness.
\end{itemize}

In Fig.~\ref{fig:homo_overview}, extremely \textit{\textbf{homogeneous}} network knowledge is shown, where each ITS-S senses a similar number of objects with similar VoI distributions. In such cases, the gain of the \textit{quality} content selector may be marginal. However, the benefit becomes more evident when the number of objects or the distribution of VoI values becomes \textit{\textbf{heterogeneous}} across ITS-S', as shown in Fig.~\ref{fig:hete_overview}.

Finally, while the example proposed in this Section interprets Figs.~\ref{fig:homo_overview} and~\ref{fig:hete_overview} across ITS-S', it may also be interpreted in temporal terms. In that reading, each row represents the content perceived by a \textit{single} ITS-S over consecutive time steps. Under \textit{homogeneous} conditions, the number of perceived objects and the VoI distribution remain broadly similar across time steps. Under \textit{heterogeneous} conditions, both may vary markedly from one time step to the next. This can also commonly occur in real world scenarios. 


\subsection{Implementation}
The operation of the proposed \textit{quality} DCC FAC approach occurs as a CPS linear rate controller which assumes the input of an allocated Facilities layer bit budget via the FCL from the ETSI \textit{Resource Management} block. In this paper, we assume this is derived from a \textit{Resource Management} congestion rate controller, conceptually similar to A-DCC, which updates a shared control variable $\delta(t)$ every control interval $T_{\mathrm{control}} = 200~\mathrm{ms}$ using the aggregate channel-load estimate (CLR). $\delta(t)$ is then converted into a service layer CPM bit rate that is used for the next two CPM generation events, with frequency $T_{\mathrm{CPM}} = 100~\mathrm{ms}$. The per-CPM application budget, $B(t)$, which is the number of bytes available for content selection at time $t$ is approximated by
\begin{equation}
\label{eq:budget_bytes}
    B(t) = \max\!\left(0,\; \delta(t)\,R\,T_{\mathrm{CPM}} - H\right)
\end{equation}
where $R$ denotes the FCL CPM radio allocation in bytes per microsecond, and $H$ denotes the fixed CPM overhead in bytes. At each CPM generation event, node $j$ computes the total size of all currently sensed content as
\begin{equation}
\label{eq:offered_load}
    L_j(t) = \sum_{i \in \mathcal{O}_j(t)} s_i^{\mathrm{B}}
\end{equation}
where $\mathcal{O}_j(t)$ is the content set at node $j$ and $s_i^{\mathrm{B}}$ is the size of a single piece of content $i$ in bytes.

Based on the current content $L_j(t)$ and $B(t)$, the ITS-S calculates its load ratio, $\rho_j(t)$, which indicates whether the current content can fit within the allocated Facilities layer bit budget:
\begin{equation}
\label{eq:load_ratio}
    \rho_j(t) =
    \begin{cases}
        \dfrac{L_j(t)}{B(t)}, & B(t) > 0\\[6pt]
        +\infty, & B(t) = 0
    \end{cases}
\end{equation}
The ITS-S then updates its local VoI threshold $\theta_j(t)$ according to Equations~\ref{eq:theta_step} and~\ref{eq:theta_update}. When $\rho_j(t) > 1$, the content set exceeds the current allocated Facilities layer bit budget and $\theta_j(t)$ increases, making the selector more restrictive. When $\rho_j(t) < 1$, the content set is below the current allocated Facilities layer bit budget and $\theta_j(t)$ decreases, allowing more content to remain eligible for CPM inclusion.
\begin{equation}
\label{eq:theta_step}
    \Delta \theta_j(t) =
    \mathrm{clip}\!\left(\beta_{\theta}\bigl(\rho_j(t)-1\bigr),\;
    \Delta \theta_{\min},\;
    \Delta \theta_{\max}\right)
\end{equation}
\begin{equation}
\label{eq:theta_update}
    \theta_j(t) =
    \mathrm{clip}\!\left((1-\alpha_{\theta})\,\theta_j(t-1) + \Delta \theta_j(t),\;
    \theta_{\min},\;
    \theta_{\max}\right)
\end{equation}
To avoid excessively minor or major upward or downward shifts in the VoI threshold under congestion, a minimum positive or negative increase/decrease ($\theta_{\min}$ or $\theta_{\max}$) can be enforced whenever the allocated Facilities layer bit budget varies.

After updating $\theta_j(t)$, the content selector discards all objects/regions whose value is below the current threshold as per Equation~\ref{eq:thresholded_set}.
\begin{equation}
\label{eq:thresholded_set}
    \mathcal{C}_j(t) = \left\{ i \in \mathcal{O}_j(t) \;:\; \mathrm{VoI}_i \geq \theta_j(t) \right\}.
\end{equation}
Hence, all content whose VoI is above the current threshold is included in the CPM. 

\section{Simulation Environment}
\label{simulation_scenario}
\begin{figure*}[t]
    \centering
    \begin{subfigure}[t]{0.48\textwidth}
        \centering
        \includegraphics[width=0.95\textwidth]{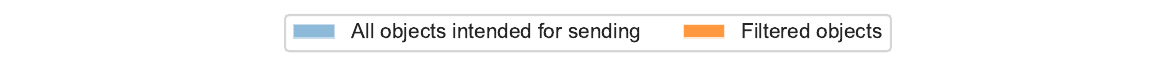}\\[-0.5mm]
        \includegraphics[width=0.98\textwidth]{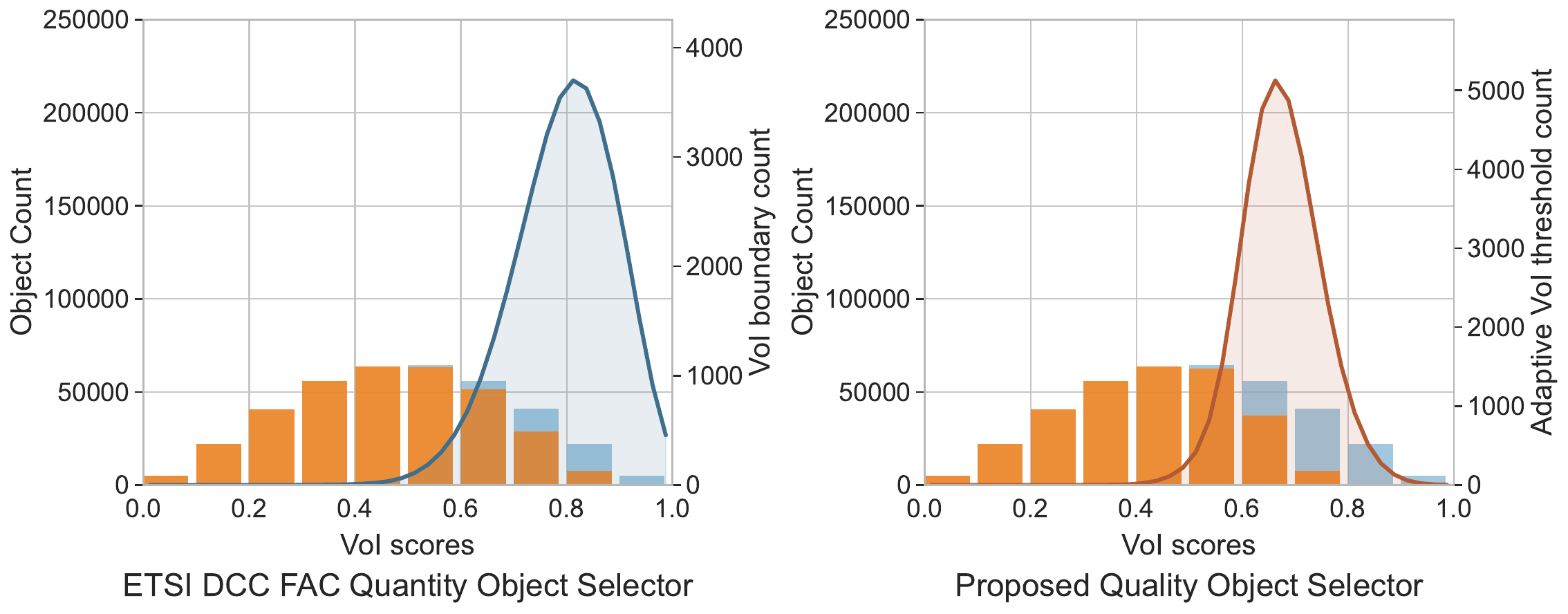}
        \caption{\textit{Homogeneous} network knowledge.}
        \label{fig:homo_voi_histograms}
    \end{subfigure}
    \hfill
    \begin{subfigure}[t]{0.48\textwidth}
        \centering
        \includegraphics[width=0.95\textwidth]{VoI_Threshold_Figures/dcc_comparison_voi_Histograms_legend.pdf}\\[-0.5mm]
        \includegraphics[width=0.98\textwidth]{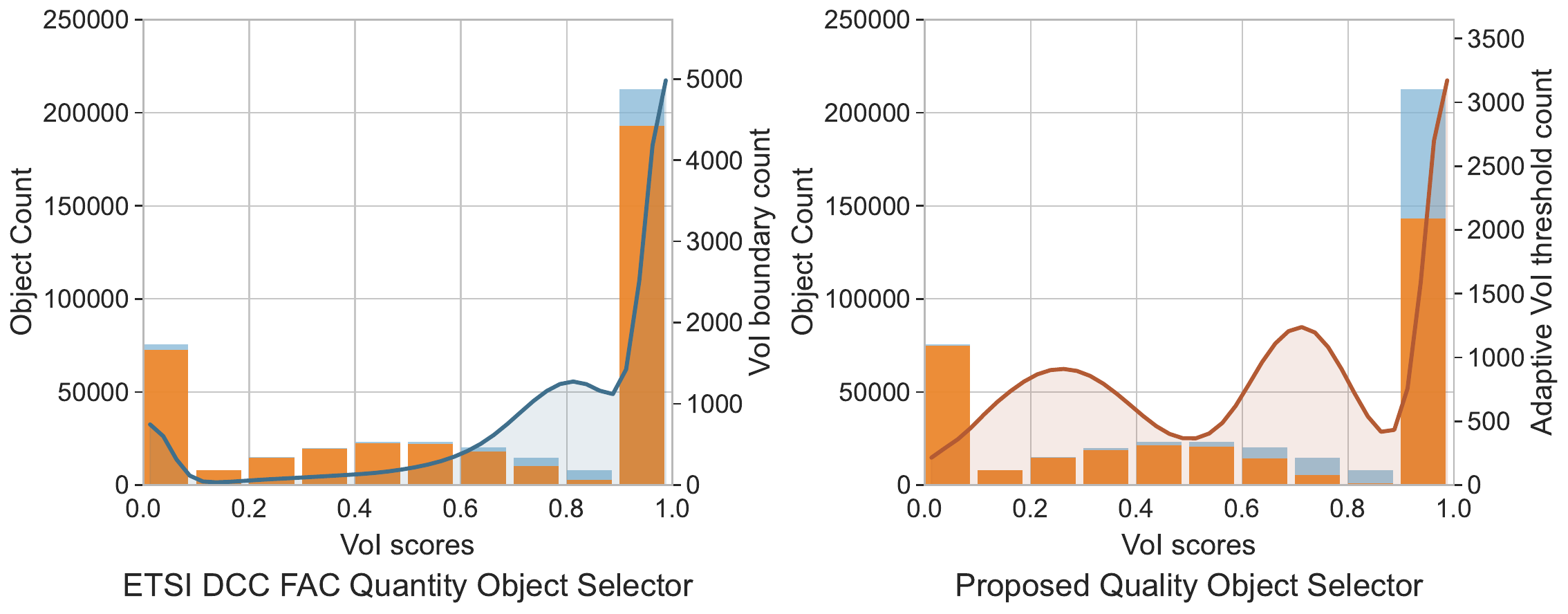}
        \caption{\textit{Heterogeneous} network knowledge.}
        \label{fig:hete_voi_histograms}
    \end{subfigure}
    \caption{Histograms of generated VoI with the objects included for transmission shown in blue. This assumes $15$ ITS-S' under (a) \textit{Homogeneous} and (b) \textit{Heterogeneous} network knowledge. The omitted $5$ and $10$ ITS-S' cases exhibit the same behaviour. Curves show the counts for the lowest VoI boundaries for the \textit{quantity} content selector and the adaptive VoI thresholds for the \textit{quality} content selectors.}
    \label{fig:voi_histograms_combined_with_theta}
    \vspace{-3mm}
\end{figure*}

\begin{table}[t]
    \centering
    \caption{Size of the content set (objects only) and distribution of VoI scores.}
    \label{tab:object_count_voi_models}
    \footnotesize
    \setlength{\tabcolsep}{3pt}
    \begin{tabular}{|l|c|c|}
        \hline
        \textbf{Profile} & \textbf{Number of Objects} & \textbf{Object's VoI} \\
        \hline
        Homogeneous & $\mu \approx 10.0,\ \sigma \approx 2.0$ & $\mu \approx 0.5,\ \sigma \approx 0.2$ \\
        \hline
        Heterogeneous: sparse & $\mu \approx 5.0,\ \sigma \approx 2.0$ & $\mu \approx 0.2,\ \sigma \approx 0.34$ \\
        \hline
        Heterogeneous: medium & $\mu \approx 10.0,\ \sigma \approx 3.0$ & $\mu \approx 0.5,\ \sigma \approx 0.2$ \\
        \hline
        Heterogeneous: dense & $\mu \approx 18.0,\ \sigma \approx 4.0$ & $\mu \approx 0.9,\ \sigma \approx 0.3$ \\
        \hline
    \end{tabular}
\end{table}

Table~\ref{tab:object_count_voi_models} summarises how homogeneous and heterogeneous network knowledge is modelled with respect to the number of objects perceived (content set size) at each CPM generation event and the associated object VoI distribution. In the \textit{homogeneous} setting, all ITS-S' share similar content set sizes and VoI distributions. In the \textit{heterogeneous} setting, ITS-S' are assigned a shuffled mix of sparse, medium, and dense profiles. Specifically, in the $5$ ITS-S' case, two ITS-S' use the sparse profile, two use the medium profile, and one uses the dense profile. In the $10$ ITS-S' case, four ITS-S' use the sparse profile, three use the medium profile, and three use the dense profile. In the $15$ ITS-S' case, five ITS-S' are assigned to each profile. Object VoI is bounded in $[0,1]$ and sampled independently from the profile-specific distribution at each CPM generation event.

The simulator abstracts the communication channel for all ITS-S'. For each 200~ms CLR update, the simulator aggregates the total transmitted CPM bytes from all ITS-S' and estimates the channel load as per~\cite{sepulcre2025v2x} according to
\begin{equation}
\label{eq:cbr_raw}
    \mathrm{CBR}_{\mathrm{raw}}(t)=\left(\frac{\mathrm{bytes}_{\mathrm{tx}}(t)}{R_{\mathrm{abs}}\,T_{\mathrm{control}}}\right)s_f
\end{equation}
where $\mathrm{bytes}_{\mathrm{tx}}(t)$ is the total number of transmitted CPM bytes within the current control window, $R_{\mathrm{abs}}$ is the abstract data rate, $T_{\mathrm{control}}=200$~ms, and $s_f$ is a scaling factor used to emulate congested operating conditions. To avoid abrupt fluctuations, the simulator applies first order smoothing:
\begin{equation}
\label{eq:cbr_smooth}
    \mathrm{CBR}(t)=\frac{1}{2}\left(\mathrm{CBR}_{\mathrm{raw}}(t)+\mathrm{CBR}(t-1)\right)
\end{equation}
The resulting CBR estimate is then used to update the global control variable $\delta(t)$, which determines the CPS FCL bit rate allocation. 

Once a set of objects has been chosen to be included in a CPM, the total message size in Bytes is approximated by
\begin{equation}   
\label{eq:cpm_abs}
\mathrm{Size}^{\mathrm{CPM}}_{\mathrm{ASN.1}} =
  \Bigg\lceil 127 + \frac{4}{8}
    + \Bigg\lceil \frac{68\pm20}{8} N_r + \frac{2}{8} \Bigg\rceil
  \Bigg\rceil 
\end{equation}
where $N_r$ is the number of included objects, the term $(68 \pm 20)$ is derived from a truncated Gaussian distribution with a mean of $68$ bits, standard deviation of $10$ bits, clipped to $[48,88]$ bits. This is used to simulate that some perceived objects are represented with richer information and therefore occupy more bits in the CPM. If no object is selected, the CPM size is set to zero and no packet is recorded as transmitted.

\section{Evaluation}
\label{evaluation}
This section compares the proposed \textit{quality} content selector against the ETSI DCC FAC \textit{quantity} content selector with results organised according to \textit{homogeneous} and \textit{heterogeneous} network knowledge. We first examine how effectively each selector preserves the \textit{network transmitted VoI} under a common allocated Facilities layer bit budget and examine the VoI threshold as network knowledge changes. We then analyse the corresponding channel-load behaviour to assess whether the VoI gains of the respective DCC approaches are achieved while maintaining stable congestion-control operation.

\subsection{Network Transmitted VoI}
\label{subsec:tx_collective_voi}
\begin{table}[t]
\centering
\caption{Network Transmitted VoI (\%) for the ETSI DCC FAC \textit{quantity} content selector and the proposed DCC FAC \textit{quality} content selector.}
\label{tab:collective_voi_retained}
\small
\setlength{\tabcolsep}{4pt}
\begin{tabular}{|l|c|c|c|}
\hline
\textbf{Setting} & \textbf{Quantity} & \textbf{Quality} & \textbf{Gain} \\ \hline
Homogenous, $5$ ITS-S'  & 11.75 & 18.28 & +6.53 \\
Homogenous, $10$ ITS-S' & 14.57 & 28.09 & +13.52 \\
Homogenous, $15$ ITS-S' & 15.51 & 32.40 & +16.89 \\ \hline
Heterogeneous, $5$ ITS-S'  &  10.06 & 23.93 & +13.87 \\
Heterogeneous, $10$ ITS-S' &  10.75 & 30.14 & +19.39 \\
Heterogeneous, $15$ ITS-S' &  10.68 & 32.00 & +21.32 \\ \hline
\end{tabular}
\vspace{-2mm}
\end{table}

This can be defined as
\begin{equation}
\label{eq:tx_collective_voi}
    \sum = \frac{\sum \mathrm{VoI}_{\mathrm{tx}}} {\sum \mathrm{VoI}_{\mathrm{gen}}}
\end{equation}
where $\sum \mathrm{VoI}_{\mathrm{tx}}$ is the total VoI of transmitted objects across all ITS-S' and $\sum \mathrm{VoI}_{\mathrm{gen}}$ is the total VoI of all generated objects over the evaluation period.
In Figs.~\ref{fig:homo_voi_histograms} and~\ref{fig:hete_voi_histograms}, the distribution of the VoI scores across all ITS-S' is shown for homogenous and heterogeneous conditions. The results are consistent with the expected behaviour of the two DCC content selectors. Under both conditions, the proposed DCC FAC \textit{quality} content selector preserves a larger fraction of the network transmitted VoI shown in blue, with the orange bars showing the content that is filtered due to bit budget limitations. It can be observed for the 15 ITS-S' case that fewer objects are excluded for transmission, especially at higher VoI scores. It confirms that given the same Facilities layer bit budget, it is beneficial to regulate object eligibility through an adaptive VoI threshold. The benefit is particularly noticeable for highly heterogeneous network knowledge as shown in Fig.~\ref{fig:hete_voi_histograms}.

A similar trend was observed across the 5 and 10 ITS-S' scenarios as shown in Table~\ref{tab:collective_voi_retained}. Considerable gains can also be observed for homogenous network knowledge across all scenarios but the gains become particularly pronounced under heterogeneous network knowledge when the size of object sets varies more significantly across ITS-S', and the VoI distribution is less uniform. In this case, gains of $13.87$, $19.39$, and $21.32$ can be observed respectively. In such a context, the ETSI DCC FAC \textit{quantity} content selector is more likely to utilise part of the available CPM budget on lower-value objects, whereas the proposed DCC FAC \textit{quality} content selector becomes more selective through the adaptive VoI threshold, thereby preserving a larger share of the most useful information.
This behaviour is reflected in Fig.~\ref{fig:hete_voi_histograms}, where the \textit{quantity} content selector retains a larger proportion of low-VoI objects, while the proposed \textit{quality} content selector shifts the transmitted set towards higher-VoI objects.

\subsection{VoI  Threshold Analysis}
To prove that the gains in network transmitted VoI are as a result of how selective the different approaches are in choosing content, we further analyse the VoI decision boundary/threshold. As the ETSI DCC quantity selector does not set a VoI threshold, the lowest VoI chosen in each CPM generation event is recorded as the "boundary" to compare against the adaptive VoI threshold of the \textit{quality} based content selector. Accordingly, both are related but not identical: the former represents the lowest transmitted VoI, whereas the latter is a true adaptive VoI control variable. 


Fig.~\ref{fig:voi_histograms_combined_with_theta} shows the overlaid VoI boundaries and thresholds for the proposed quantity and \textit{quality} content selectors, respectively. Under \textit{heterogeneous} network knowledge in Fig.~\ref{fig:hete_voi_histograms}, the distinction becomes apparent. The proposed \textit{quality} content selector clearly exhibits a multi-modal threshold distribution, with three visible peaks that align closely with the sparse, medium, and dense ITS-S profiles defined in the simulation. This demonstrates that the object selector responds directly to profile-dependent sensing and load conditions, producing adapted VoI thresholds rather than a single undifferentiated boundary based on an allocated bit budget. In other words, the selector applies distinct decision boundaries to ITS-S' with different object characteristics, which is precisely the intended behaviour in a heterogeneous setting. In contrast, the \textit{quantity} content selector remains concentrated towards high VoI values, with a smaller lower-valued tail, indicating that the effective boundary inferred from transmitted objects is dominated by the strongest retained objects and therefore fails to capture the underlying heterogeneity explicitly. The lower-valued tail also shows that the \textit{quantity} content selector still admits low-VoI objects in some CPM generation events, whereas the adaptive threshold of the proposed \textit{quality} content selector better separates low, medium, and high demand operating conditions, reducing the likelihood that low-VoI objects consume the allocated CPM budget when higher-value content is present.

For \textit{homogeneous} network knowledge shown in Fig.~\ref{fig:homo_voi_histograms}, the VoI threshold for the proposed \textit{quality} content selector is more targeted than the broader VoI boundaries observed for the \textit{quantity} selector. This indicates that the adaptive VoI threshold converges when all ITS-S' observe statistically similar object sets. This is consistent with the more selective exclusion of low VoI objects, such that the retained distribution remains concentrated towards the upper VoI range. Although the boundaries of the \textit{quantity} selector are centred at a higher VoI level, this should not be interpreted as evidence of a stricter or more effective thresholding mechanism; rather, it reflects the fact that the plotted values are the weakest VoIs that were still transmitted in each CPM. 

\subsection{Channel Busy Ratio}
\label{subsec:cbr}
\begin{figure*}[t]
    \centering
    \begin{subfigure}[t]{0.48\textwidth}
        \centering        \includegraphics[width=0.62\textwidth]{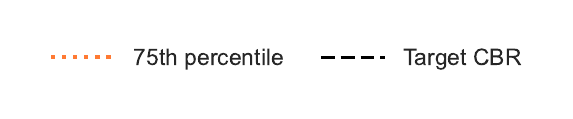}\\[-1.8mm]
        \includegraphics[width=0.90\textwidth]{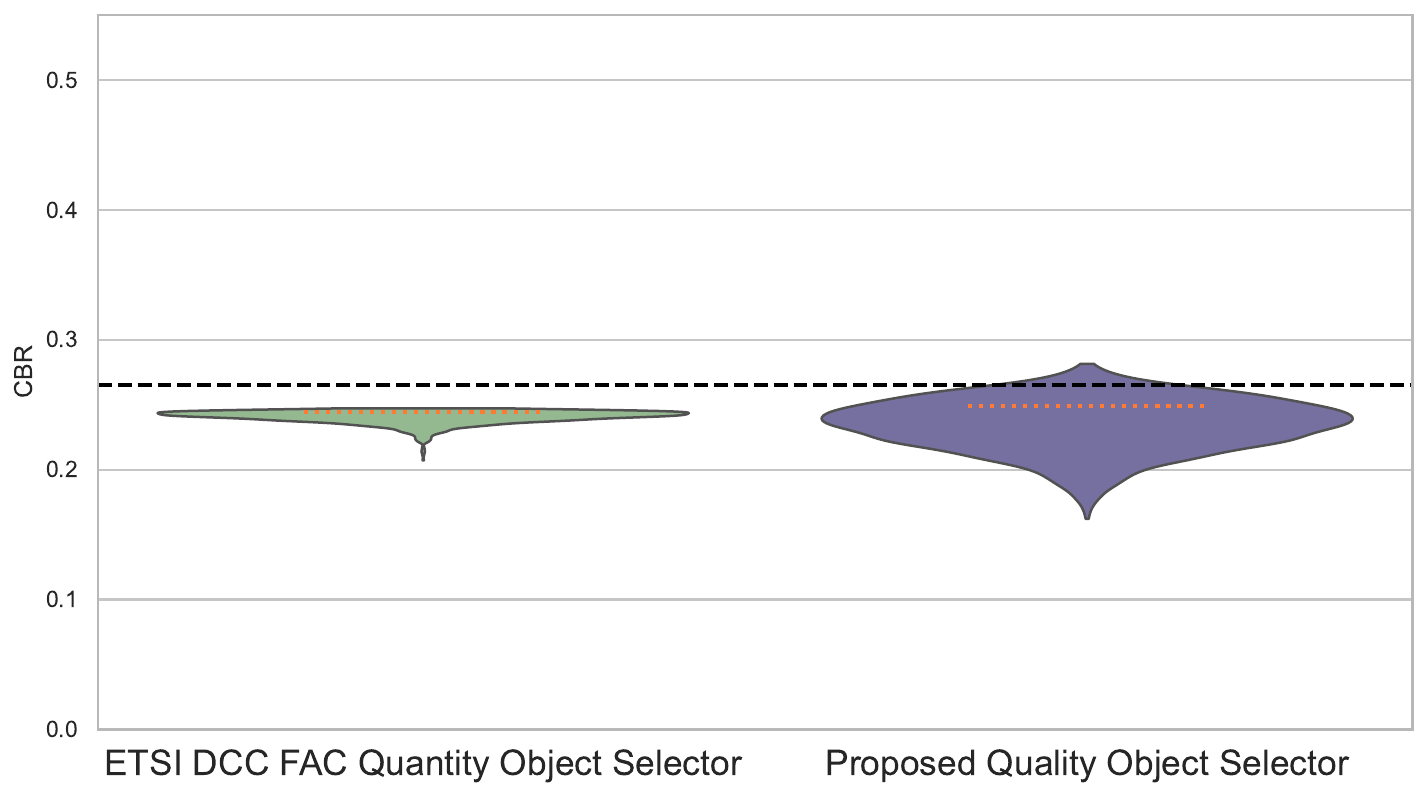}
        \caption{\textit{Homogeneous} network knowledge.}
        \label{fig:homo_cbr_violins}
    \end{subfigure}
    \hfill
    \begin{subfigure}[t]{0.48\textwidth}
        \centering
        \includegraphics[width=0.62\textwidth]{VoI_Threshold_Figures/cbr_violin_legend.pdf}\\[-1.0mm]
        \includegraphics[width=0.90\textwidth]{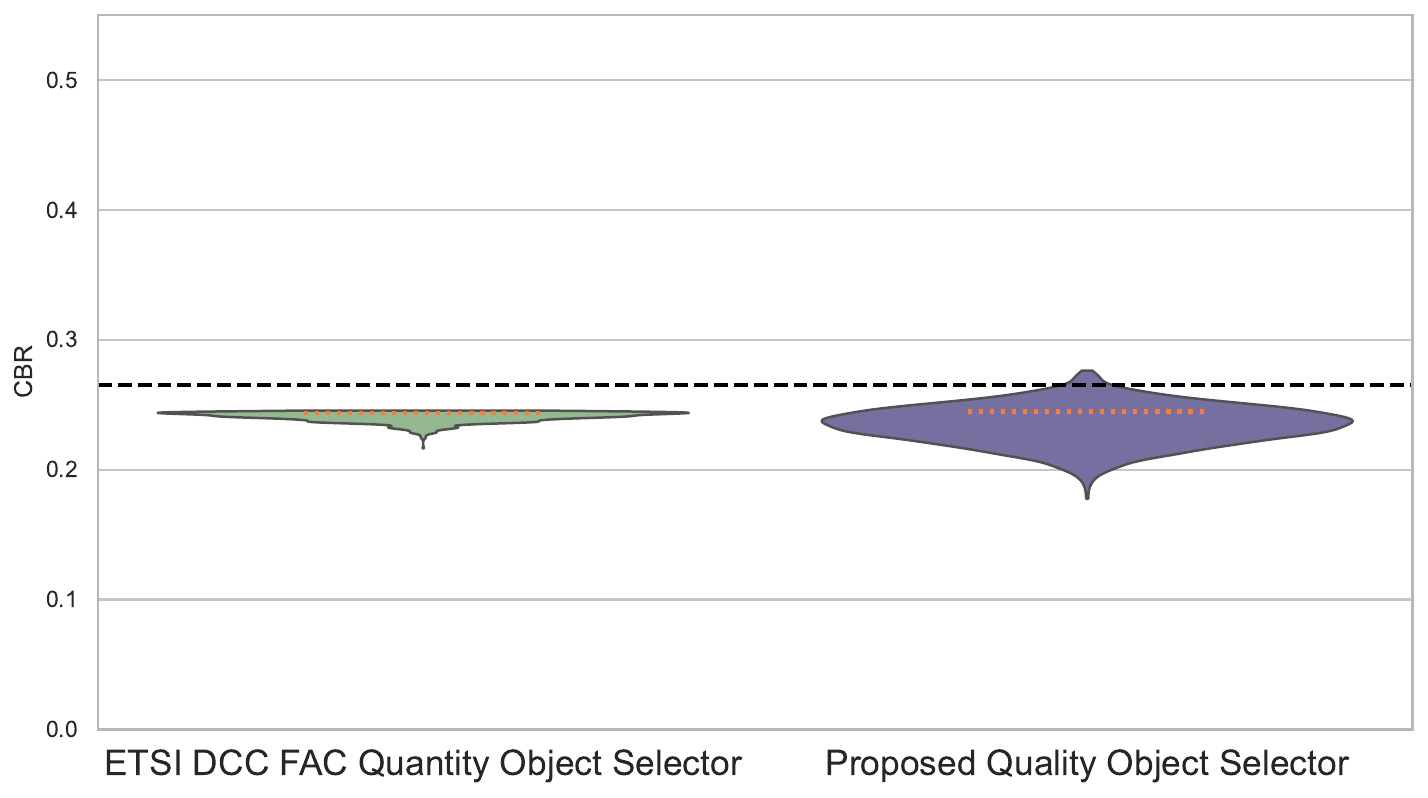}
        \caption{\textit{Heterogeneous} network knowledge.}
        \label{fig:hete_cbr_violins}
    \end{subfigure}
    \caption{Representative CBR violin plots for the $15$ ITS-S' case under (a) \textit{homogeneous} and (b) \textit{heterogeneous} network knowledge. The dashed line indicates the target CBR. The omitted $5$ and $10$ ITS-S' cases exhibit the same behaviour.}
    \label{fig:cbr_violins_combined}
    \vspace{-3mm}
\end{figure*}

Fig.~\ref{fig:cbr_violins_combined} compares representative CBR for both content selectors. CBR is maintained close to the target level in both cases, indicating that they are capable of controlling congestion. This is important because it shows that the improvement in \textit{network transmitted VoI} reported by the \textit{quality} content selector is not achieved by disregarding channel stability. However an under-, and occasional over-utilisation of the channel can occur. This is not a limitation of the adaptive VoI threshold, but rather arises from how the channel resources are allocated by the \textit{BME} shown in the reference architecture in Fig.~\ref{fig:etsi_dcc_fac}. The BME allocates the Facilities layer bit budget to the respective service, and the CPS refines usage by adapting the VoI threshold accordingly. The reference architecture does not state how the BME should allocate resources. In this paper, an A-DCC type rate controller allocates resources \textit{equally} across ITS-S'. If too few resources are allocated the VoI threshold increases and if too many are allocated, it decreases. 

However there is no feedback mechanism to the BME rate controller to preferentially allocate a higher Facilities layer bit budget to the ITS-S' with high value perception data. This could be facilitated via the FCP in Fig.~\ref{fig:etsi_dcc_fac}. Only one recent paper has attempted to address this~\cite{sepulcre2026demand} by specifying a BME rate controller that weights allocated Facilities layer bit budget based on individual service demands in the FCP and the service priority. They assume the VoI distribution increases and decreases linearly on a given ITS-S'. While promising, this approach remains content-independent, i.e. it reflects demand mainly through the \textit{quantity} of data demand rather than the \textit{value} of the sensed perception data. Thus it is important to investigate \textit{VoI-based demand allocation} at the BME, so that service-level allocated Facilities layer bit budget can better reflect not only channel load and traffic demand, but also the instantaneous value of the perception data offered by the CPS. Designing a non-linear BME rate controller that includes a feedback based control loop will be the focus of future work.

\section{Conclusions \& Future Work}
\label{conclusions}
This paper examined the limitations of current DCC mechanisms in prioritising the transmission of collectively perceived high-value data while maintaining a target channel load. In particular, it was shown that the ETSI DCC FAC \textit{quantity} content selector, although representing an important first step towards content awareness, does not maximise \textit{network transmitted VoI} under heterogeneous network knowledge, where the number of sensed objects and their VoI distributions vary more significantly across ITS-S' and over time. To address this, this paper proposed a DCC FAC \textit{quality} content selector that dynamically adapts the content set included in each CPM through an adaptive VoI threshold. The evaluation showed that the proposed method produces gains in the \textit{network transmitted VoI} under both homogeneous and heterogeneous network knowledge while maintaining channel load close to the target CBR. 

Several directions remain for future work. First, the proposed approach should be validated using real world datasets together with richer sensor perception models and enhanced communication channel modelling. Second, the channel-load control and convergence behaviour of the BME rate controller deserves further study, particularly with respect to the interaction with the adaptive VoI-threshold mechanism. Third, this study assumes a fixed CPM generation interval of $100$~ms. In a practical CPS, however, the CPM generation interval should be adapted according to the VoI distribution of the content. 
Such an adaptive interval would directly interact with the adaptive VoI threshold, since changing the CPM generation rate also changes the available transmission opportunity and the instantaneous content-selection pressure. The specific formulation and implementation of this interval-adaptation function is left for future work.   

\section*{Acknowledgment}
This publication has emanated from research conducted with the financial support of Taighde Éireann - Research Ireland under Grant numbers 18/CRT/6222 \& 13/RC/2077 P2. For the purpose of Open Access, the author has applied a CC BY public copyright licence to any Author Accepted Manuscript version arising from this submission.

\bibliography{references}
\bibliographystyle{IEEEtran}
\end{document}